# Time weakens the Bell's inequalities.

Alejandro A. Hnilo.

CEILAP, Centro de Investigaciones en Láseres y Aplicaciones, UNIDEF(MINDEF-CONICET).

CITEDEF, J.B. de La Salle 4397, (1603) Villa Martelli, Argentina.

*emails*: ahnilo@citedef.gob.ar, alex.hnilo@gmail.com

September 26th, 2013.

*Abstract*

*By taking into account that all real measurements are performed successively, during time, it is concluded that the violation of the Bell's inequalities in the Nature does not refute (even in an ideally perfect experiment) the theories holding to Local Realism, for an unavoidable additional assumption is involved. Yet, in order to be acceptable, such theories must predict different values for the factual and counterfactual time averages of probabilities or observables.*

PACS: 03.65.Ud Entanglement and quantum non-locality (EPR paradox, Bell's inequalities, etc.) - 03.65.Ta Foundations of quantum mechanics - 42.50.Xa Optical tests of quantum theory.

**1. Introduction.**

The Bell's inequalities [1] are correlation bounds inferred from the innate ideas of Locality (roughly speaking: the results of a measurement are independent of the events outside its past lightcone) and Realism (the properties of the physical world are well defined, regardless of whether it is being observed or not). Some predictions of Quantum Mechanics (QM) violate those bounds, so that QM is concluded to be incompatible with Local Realism (LR). This contradiction is a crucial issue, because LR is assumed not only in everyday life, but also in all the scientific practice (excepting QM).

For almost 50 years, the experiments devoted to test whether the Bell's inequalities are violated in the Nature, or not, have been the touchstone to settle the controversy between QM and LR. Many researchers (including this author) have made their best efforts to perform experiments under increasingly stringent conditions, to close any doubt on the results that may arise from practical imperfections (the so-called *logical loopholes*). The strength of the Bell's inequalities is that they take into account *all the possible ways* in which a classical system can produce correlated results between measurements performed on two remote particles. In order to produce such correlations, each particle is supposed to carry a "hidden variable". A key element in the construction of the Bell's inequalities is that there is a complete freedom in the choice of the hidden variable, as far as LR is not violated.

In this paper, it is shown that, in order to apply the Bell's inequalities to a real measuring process, it is necessary to make *at least one* assumption additional to LR, even in an ideally perfect setup. The key of the analysis leading to this result is that all real measurements are situated in time. The necessity of the additional assumption arises from the logical impossibility of measuring with two different macroscopic settings at the same time. In all the conceivable cases, the additional assumption is far less deep-rooted than LR. Therefore, it is sensible to interpret the observed violation of the Bell's inequalities as a refutation of that additional assumption, and not of LR. Moreover, depending on the particular additional assumption made, the QM predictions violate the bound, or not. In consequence, no experiment based on testing the violation of the Bell's inequalities can be decisive in the QM vs LR controversy. Nevertheless, the set of LR theories able to violate the Bell's inequalities is restricted in a precise way, which is explained in the text and provides a useful tool for further research.

In the next Section, the usual reasoning leading to two of the main forms of the Bell's inequalities (Clauser-Horne, CH [2] and Clauser-Horne-Shimony and Holt, CHSH [3]) is briefly reviewed. The Reader familiar with them may go directly to the next Section 3, where it is shown how these inequalities are affected if time is taken into account. The consequences of different additional assumptions are presented in the Section 4. The results are discussed in the Section 5.

**2. Usual derivation of the CH and CHSH inequalities.**

Assume that the probability to detect a photon after an analyzer A oriented at an angle $\alpha$ (see Fig.1) is $P_A(\alpha,\lambda)$, where $\lambda$ is an arbitrary "hidden" variable. The strength of the Bell's inequalities is that $\lambda$ can be anything: it can be real or complex, a vector, a tensor, etc. The only necessary assumption is that the integrals over the $\lambda$-space exist. The observable probability of detection is:

$$P_A(\alpha) = \int d\lambda.\rho(\lambda). P_A(\alpha,\lambda) \qquad (1)$$

where $\rho(\lambda)$ is a normalized distribution in the $\lambda$-space ($\int d\lambda.\rho(\lambda)=1$). These assumptions are in compliance with *Realism*. Consider now two photons carrying the same value of $\lambda$. The probability that both photons are detected after analyzers A and B set at angles $\{\alpha,\beta\}$ is, by definition, $P_{AB}(\alpha,\beta,\lambda)$. *Locality* implies that:

$$P_{AB}(\alpha,\beta,\lambda) = P_A(\alpha,\lambda) \times P_B(\beta,\lambda). \qquad (2)$$

so that the probability to observe a double detection is:

$$P_{AB}(\alpha,\beta) = \int d\lambda \cdot \rho(\lambda) \cdot P_A(\alpha,\lambda) \cdot P_B(\beta,\lambda). \tag{3}$$

Locality also implies that $\{\alpha,\beta,\lambda\}$ are statistically independent variables: $P_A(\alpha,\lambda) = P_A(\alpha) \times P_A(\lambda)$, and that $\rho(\lambda)$ is independent of $\{\alpha,\beta\}$. These properties often receive the specific name of *measurement independence*. To enforce Locality and measurement independence in the practice, notable experiments have been performed [4,5] with $\{\alpha,\beta\}$ changing in a time shorter than $L/c$. In what follows, Locality and measurement independence are taken for granted.

Given $\{x,y \geq 0, X \geq x', Y \geq y'\}$ the following equality holds:

$$-1 \leq xy - xy' + x'y + x'y' - Xy - Yx' \leq 0 \tag{4}$$

choosing $x = P_A(\alpha,\lambda)$, $x' = P_A(\alpha',\lambda)$, $y = P_B(\beta,\lambda)$, $y' = P_B(\beta',\lambda)$ and $X=Y=1$:

$$-1 \leq P_A(\alpha,\lambda) \cdot P_B(\beta,\lambda) - P_A(\alpha,\lambda) \cdot P_B(\beta',\lambda) + P_A(\alpha',\lambda) \cdot P_B(\beta,\lambda) +$$
$$+ P_A(\alpha',\lambda) \cdot P_B(\beta',\lambda) - P_B(\beta,\lambda) - P_A(\alpha',\lambda) \leq 0 \tag{5}$$

where $\{\alpha,\beta,\alpha',\beta'\}$ are different analyzers' orientations. Applying $\int d\lambda \cdot \rho(\lambda)$ and eq.3:

$$-1 \leq P_{AB}(\alpha,\beta) - P_{AB}(\alpha,\beta') + P_{AB}(\alpha',\beta) + P_{AB}(\alpha',\beta') - P_B(\beta) - P_A(\alpha') \equiv M \leq 0 \tag{6}$$

Each single photon is observed non-polarized, then $P_A(\alpha) = P_A(\beta) = \frac{1}{2}$ so that:

$$0 \leq P_{AB}(\alpha,\beta) - P_{AB}(\alpha,\beta') + P_{AB}(\alpha',\beta) + P_{AB}(\alpha',\beta') \leq 1 \tag{7}$$

which is the CH inequality. The QM predictions violate it. F.ex., for the entangled state $|\phi^+\rangle = (1/\sqrt{2})\{|x_a,x_b\rangle + |y_a,y_b\rangle\}$, $P_{AB}(\alpha,\beta) = \frac{1}{2}\cos^2(\alpha,\beta)$, and choosing $\{0, \pi/8, \pi/4, 3\pi/8\}$ as $\{\alpha,\beta,\alpha',\beta'\}$:

$$0 \leq 0.427 - 0.073 + 0.427 + 0.427 = \frac{1}{2}(1+\sqrt{2}) = 1.207 \leq 1 \tag{8}$$

that violates the inequality. Therefore, QM is incompatible with at least one of the assumptions leading to eq.7 (i.e., Locality and/or Realism).

The CHSH inequality involves, instead of probabilities, the expectation values of observables. Let define the observable result +1 (-1) if the photon is detected in the transmitted (reflected) port of the analyzers in Fig.1. The observable corresponding to a coincident detection $AB(\alpha,\beta,\lambda) \equiv A(\alpha,\lambda) \times B(\beta,\lambda)$ is:

$$AB(\alpha,\beta,\lambda) = \{(+1).[C^{++} + C^{--}] + (-1)[C^{-+} + C^{+-}]\}/\{C^{++} + C^{--} + C^{-+} + C^{+-}\}|_{\alpha,\beta,\lambda} \qquad (9)$$

where $C^{ij}$ is the number of coincident photons detected with results "$i$" and "$j$" ($i= \pm 1, j= \pm 1$) for the angle settings ($\alpha,\beta$) and value $\lambda$ of the hidden variable. Its expectation value is then the average over the $\lambda$-space:

$$E(\alpha,\beta) = \int d\lambda.\rho(\lambda).AB(\alpha,\beta,\lambda). \qquad (10)$$

hence, for two different angle settings $\{\alpha,\beta\}$ and $\{\alpha,\beta'\}$:

$$E(\alpha,\beta) - E(\alpha,\beta') = \int d\lambda.\rho(\lambda).AB(\alpha,\beta,\lambda) - \int d\lambda.\rho(\lambda).AB(\alpha,\beta',\lambda) = \qquad (11)$$
$$= \int d\lambda.\rho(\lambda).[AB(\alpha,\beta,\lambda) - AB(\alpha,\beta',\lambda)]$$

After adding and subtracting the term $[AB(\alpha,\beta,\lambda).AB(\alpha',\beta,\lambda).AB(\alpha,\beta',\lambda).AB(\alpha',\beta',\lambda)]$ inside the integral, reordering and applying the modulus (note that $1\pm AB \geq 0$):

$$|E(\alpha,\beta) - E(\alpha,\beta')| \leq \int d\lambda.\rho(\lambda).[1 \pm AB(\alpha',\beta',\lambda)] + \int d\lambda.\rho(\lambda).[1 \pm AB(\alpha',\beta,\lambda)] = \qquad (12)$$
$$= 2 \pm [E(\alpha',\beta') + E(\alpha',\beta)]$$

and the final form of the CHSH inequality is:

$$S_{CHSH} \equiv |E(\alpha,\beta) - E(\alpha,\beta')| + |E(\alpha',\beta') + E(\alpha',\beta)| \leq 2 \qquad (13)$$

For the same entangled state and angle settings than for eq.8, the QM prediction for the lhs of eq.13 is $2\sqrt{2}$, violating the bound.

**3. Modification of the CH and CHSH inequalities if time is taken into account.**

All real measurements occur successively, during time. The expression for the observable probabilities are then (f.ex., for $P_A$ in eq.1):

$$P_A(\alpha) = (1/\Delta t) \int_{\theta}^{\theta+\Delta t} dt.\rho(t).P_A(\alpha,t) \qquad (14)$$

This equation represents the result of the following real process: set A=α during the time interval [$\theta, \theta+\Delta t$], sum up the number of photons detected after the analyzer, and obtain $P_A(\alpha)$ as the ratio of detected over incident photons. Note that eq.14 is equivalent to eq.1 if time is chosen as the hidden variable. Following the usual reasoning leading to the CH inequality, let apply now $(1/T)\int_0^T dt.\rho(t)$ in eq.5, where $T$ is the total measuring time:

$$-T \leq \int_0^T dt.\rho(t).P_{AB}(\alpha,\beta,t) - \int_0^T dt.\rho(t).P_{AB}(\alpha,\beta',t) + \int_0^T dt.\rho(t).P_{AB}(\alpha',\beta,t) + \int_0^T dt.\rho(t).P_{AB}(\alpha',\beta',t) -$$

$$- \int_0^T dt.\rho(t).P_B(\beta,t) - \int_0^T dt.\rho(t).P_A(\alpha',t) \leq 0 \qquad (15)$$

But, eq.15 does not correspond to what is actually measured. In most experiments on the Bell's inequalities, the measuring time is distributed in a way similar to the following: the analyzer A is set to α between t=0 and t=T/2 and to α' between t=T/2 and t=T; B=β between t=T/4 and t=3T/4, and B=β' between t=0 and t=T/4, and also between t=3T/4 and t=T (Fig.2). A different distribution requires a more involved notation of the integration intervals, but the result is the same, for Locality and measurement independence are assumed valid. The value of M (see eq.6) that is actually measured in an experiment can be written then as:

$$(1/\Delta T)\int_{T/4}^{T/2} dt_2.\rho(t_2).P_{AB}(\alpha,\beta,t_2) - (1/\Delta T)\int_0^{T/4} dt_1.\rho(t_1).P_{AB}(\alpha,\beta',t_1) + (1/\Delta T)\int_{T/2}^{3T/4} dt_3.\rho(t_3).P_{AB}(\alpha',\beta,t_3) +$$

$$+ (1/\Delta T)\int_{3T/4}^T dt_4.\rho(t_4).P_{AB}(\alpha',\beta',t_4) - (1/2\Delta T)\int_{T/4}^{3T/4} dt".\rho(t").P_B(\beta,t") - (1/2\Delta T)\int_{T/2}^T dt'.\rho(t').P_A(\alpha',t')$$

$$(16)$$

where $\Delta T = \tfrac{1}{4}T$ and the integration variable is indicated with different names in each term in benefit of clarity. Note that all the integrals in the eqs.15 and 16 are over different intervals of integration. Therefore, the eq.16 is not necessarily bounded by the inequality in eq.15. The step from eq.15 to 16, which is obvious (from eq.5 to 6) in the usual CH reasoning, is not immediate now. The values measured in eq.16 may violate the CH inequality or not but, at this point, this result implies nothing about LR. To restore the logical link to the LR hypotheses, it is necessary to bridge the gap between

eqs.15 and 16. The usual way to do it is to take for granted that (f.ex., for the last term in both equations):

$$(1/T)\int_0^T dt.\rho(t).P_A(\alpha',t) = (1/2\Delta T)\int_{T/2}^T dt'.\rho(t').P_A(\alpha',t') \qquad (17)$$

and analogous equalities for the other terms. According to the eq.14, the eq.17 holds if the analyzer is set to the same orientation during [0,*T*] and [*T*/2,*T*]. But, let see the lhs of eq.17 in detail:

$$(1/T)\int_0^T dt.\rho(t).P_A(\alpha',t) = (1/T)\int_0^{T/2} dt.\rho(t).\underline{\underline{P_A(\alpha',t)}} + (1/T)\int_{T/2}^T dt'.\rho(t').P_A(\alpha',t') \qquad (18)$$

using eq.14, the second term in the rhs is calculated equal to ¼ (note the importance of using the correct normalization time). The first term, instead, is impossible to calculate, for the underlined factor means: "the probability that a photon passes the analyzer in A if A=α', *when* A=α (≠α')". The eq.14 cannot be applied to calculate the first term. The "intuitive answer" is that $P_A(\alpha',t)=0$ (see the consequences in the Section 4.C), but this is not necessarily true. Note that:

$$P_A(\alpha',t) = \text{Probability}(\{pass\}\cap\{A=\alpha'\}\cap\{0\leq t\leq T/2\}) / \text{Probability}(\{A=\alpha'\}\cap\{0\leq t\leq T/2\}) \qquad (19)$$

but {A=α'}∩{0≤t≤*T*/2}=∅, because A=α when 0 ≤ t ≤ *T*/2. The eq.19 is a "zero-over-zero" indeterminacy: $P_A(\alpha',t)$ can take any value. This is because $P_A(\alpha',t)$ here is the probability of an event under conditions that do not occur; it is a counterfactual probability. Similar counterfactual probabilities appear in all the integrals in eq.15. F.ex., in the first integral:

$$\int_0^T dt.\rho(t).P_{AB}(\alpha,\beta,t) = \int_0^{T/4} dt.\rho(t).P_A(\alpha,t).\underline{\underline{P_B(\beta,t)}} + \int_{T/4}^{T/2} dt.\rho(t).P_A(\alpha,t).P_B(\beta,t) +$$

$$+ \int_{T/2}^{3T/4} dt.\rho(t).\underline{\underline{P_A(\alpha,t)}}.P_B(\beta,t) + \int_{3T/4}^T dt.\rho(t).\underline{\underline{P_A(\alpha,t).P_B(\beta,t)}} \qquad (20)$$

the underlined factors indicate the counterfactual probabilities, whose values are unknown.

If counterfactual reasoning is rejected, the logical gap between eq.15 and 16 remains irremediably open, and the Bell's inequalities cannot be applied to measurements performed in time, i.e., to any real measurement. It is curious that rejecting counterfactual reasoning (what is usually mandatory in QM) blocks the applicability of the main experimental support of QM in the QM vs LR controversy.

From the point of view of Realism, instead, counterfactual reasoning is natural, but a "possible world" must be defined, in addition to LR, to ensure logical consistency and to assign numerical values to the counterfactual magnitudes [6]. As it is shown in the next Sections, depending of the possible world chosen, the CH and CHSH inequalities take a different form.

There is no mystery in this situation, but simply lack of information. Let see an example of everyday life: if when I go to the cafeteria I have 30% probability of finding my friend Alice there, then: "what is the probability for me to find Alice in the cafeteria *when I don't go there*?" If the question is strictly considered, there is no answer. If it is assumed that Alice and the cafeteria have a well defined existence even when I do not go there (roughly speaking, if Realism is assumed), the question can be rephrased as: "what is the probability for *someone* to find Alice in the cafeteria when I don't go there?" which does have a well defined answer, say, $q$. But the only available information at this point is that the probability I find Alice is 30%, so that the numerical value of $q$ is unknown. More information is needed (i.e.: a possible world must be defined) to assign a numerical value to $q$.

Once a possible world is defined, the underlined factors in eq.18 and 20 (and in all the other integrals in eq.15) can be calculated in a consistent way, and the gap between eqs.15 and 16 is bridged. Yet, the definition of a possible world unavoidably involves one assumption *additional* to LR, thus weakening the consequences of the observed violation of the Bell's inequalities. Note that this weakening does not arise from an experimental imperfection (as in the case of the logical loopholes). The setup is assumed ideally perfect. The weakening arises only from the fact that real measurements are performed during time, and that it is impossible to measure with two different angle settings at the same time.

The CHSH inequality is affected in the same way. The critical step is the apparently innocent passage from the first to the second line of eq.11, which actually is:

$$E(\alpha,\beta) - E(\alpha,\beta') = (1/\Delta T)\int_{T/4}^{T/2} dt.\rho(t).AB(\alpha,\beta,t) - (1/\Delta T)\int_{0}^{T/4} dt.\rho(t).AB(\alpha,\beta',t)$$

$$\neq (1/T)\int_{0}^{T} dt.\rho(t).[AB(\alpha,\beta,t) - AB(\alpha,\beta',t)] \qquad (21)$$

The rhs in the first line is what is measured, while the integral in the second line is the expression that leads to the CHSH inequality. In order to retrieve the usual reasoning, one must add counterfactual expectation values $\underline{\underline{E}}$ to the lhs and take into account the correct normalization factor in the rhs:

$$E(\alpha,\beta)+\underline{\underline{E(\alpha,\beta)}}-E(\alpha,\beta')-\underline{\underline{E(\alpha,\beta')}}=(1/\Delta T)\int_0^T dt.\rho(t).[AB(\alpha,\beta,t)-AB(\alpha,\beta',t)] \qquad (22)$$

where:

$$\underline{\underline{E(\alpha,\beta)}}=(1/\Delta T)\int_0^{T/4} dt.\rho(t).A\underline{\underline{B}}(\alpha,\beta,t)+(1/\Delta T)\int_{T/2}^{3T/4} dt.\rho(t).\underline{\underline{AB}}(\alpha,\beta,t)+(1/\Delta T)\int_{3T/4}^{T} dt.\rho(t).\underline{\underline{AB}}(\alpha,\beta,t) \qquad (23)$$

$$\underline{\underline{E(\alpha,\beta')}}=(1/\Delta T)\int_{T/4}^{T/2} dt.\rho(t).A\underline{\underline{B}}(\alpha,\beta',t)+(1/\Delta T)\int_{T/2}^{3T/4} dt.\rho(t).\underline{\underline{AB}}(\alpha,\beta',t)+(1/\Delta T)\int_{3T/4}^{T} dt.\rho(t).A\underline{\underline{B}}(\alpha,\beta',t) \qquad (24)$$

where the (f.ex.) factor $A\underline{B}(\alpha,\beta,t)$ indicates the result of a measurement performed at a time value when B≠β, i.e., a counterfactual result. Note that there are terms with double counterfactuals. Assuming that the counterfactual results can only take the values ±1, the reasoning follows as it is usual, but the final expression is:

$$\left|E(\alpha,\beta)+\underline{\underline{E(\alpha,\beta)}}-E(\alpha,\beta')-\underline{\underline{E(\alpha,\beta')}}\right|+\left|E(\alpha',\beta')+\underline{\underline{E(\alpha',\beta')}}+E(\alpha',\beta)+\underline{\underline{E(\alpha',\beta)}}\right|\leq 8 \qquad (25)$$

I stress that it is this, and not eq.13, the inequality that is inferred from LR *only*. Note that the bound is not as trivial as it may appear, for the counterfactual expectation values are bounded by ±3 (not by ±1). From eq.9, the (f.ex.) observable $A\underline{B}(\alpha,\beta,t)$ is:

$$A\underline{B}(\alpha,\beta,t) = \{(+1).[C^{++}+C^{--}]+(-1)[C^{-+}+C^{+-}]\}/\{C^{++}+C^{--}+C^{-+}+C^{+-}\}|_{\alpha,\beta,0<t<T/4} \qquad (26)$$

whose value is not determined, because all the $C^{ij}$=0. As in eq.19, it is a zero-over-zero indeterminacy. Once again, this is a counterfactual situation that can be escaped only by defining a possible world. In other words: Realism allows the logical inference of the CHSH inequality eq.25, but Realism alone does *not* suffice to give numerical values to the counterfactual terms. One must make an additional hypothesis (i.e., to define a "possible world") to be able to calculate the value of the indeterminacy in eq.26, and hence the counterfactual terms in eq.25.

In the following Section, the meaning and consequences of four possible worlds are discussed, for both the CH and the CHSH inequalities.

## 4. Four "possible worlds".

*4.A The factual and counterfactual probabilities (observables in CHSH) are equal.*

That is, $P_A(\alpha',t)=P_A(\alpha',t')$ in eq.17 or $\underline{\underline{AB}}(\alpha,\beta,t') = AB(\alpha,\beta,t)$ in eq.23 (and the same for all the other counterfactuals), and then the usual CH and CHSH inequalities are immediately retrieved. This is the possible world implicit in the Eberhard inequality [7], which is a form of the CH inequality.

The equalities above are assumed to hold for all the counterfactuals and time values, so that the probabilities (or the expectation values) become, in the practice, independent of time. As time here plays the role of the hidden variable, this possible world has the same consequences than assuming that there is no hidden variable at all. This possible world is thus not convenient, because one is *assuming* the result that one would like to be able to *demonstrate* (i.e.: that there are no hidden variables). Actually, it is even worse. The logic inference becomes:

*LR + there are no hidden variables $\Rightarrow$ Bell's inequalities are valid.*

Therefore, the observed violation of the Bell's inequalities in the Nature might be used *in support* of the existence of hidden variables, what is the opposite of the usual interpretation. A less restrictive and, in my opinion, more meaningful possible world is the one that follows:

*4.B The factual and counterfactual time averaged probabilities (expectation values in CHSH) are equal.*

This means that eq.17 is valid, and the same for all the other counterfactual integrals. The usual CH inequality is then retrieved. In the case of CHSH, each of the three counterfactual terms in the rhs in eqs.23 and 24 becomes equal to the factual term, therefore:

$$\underline{\underline{E(\alpha,\beta)}} = 3 \times E(\alpha,\beta) \qquad (27)$$

and the usual CHSH inequality is retrieved too (see eq.25).

In my opinion, it is under this assumption (inadvertently made) that the Bell's inequalities have been applied to interpret the experimental results. At this point, the important consequence is that the observed violation of the inequalities in the Nature does not necessarily imply that LR is false: the assumed equality between the factual and the counterfactual time averages or expectation values may be false instead. Faced to choose between LR and eq.17 (or eq.27), I find sensible to pick the latter as false. Unless a solid argument can be presented to justify the validity of eqs.17 and 27 (and the argument should be notoriously *more solid* than LR, in all senses) it must be concluded that the observed violation of the Bell's inequalities demonstrates that the time averages of factual and counterfactual probabilities (or expectation values) are different. Note that there is no direct way to measure the time average of a counterfactual probability or expectation value.

*4.C. The counterfactual probabilities (expectation values in CHSH) are zero.*

This is the "intuitive answer" mentioned before. Choosing then $P_A(\alpha',t) = 0$ (and the same for all the other counterfactual probabilities) the eq.6 becomes:

$$-1 \leq \tfrac{1}{4} P_{AB}(\alpha,\beta) - \tfrac{1}{4} P_{AB}(\alpha,\beta') + \tfrac{1}{4} P_{AB}(\alpha',\beta) + \tfrac{1}{4} P_{AB}(\alpha',\beta') - \tfrac{1}{2} P_B(\beta) - \tfrac{1}{2} P_A(\alpha') \leq 0 \qquad (28)$$

using the QM values of the probabilities for the usual choice of angle settings, the numbers are:

$$-1 \leq \tfrac{1}{4} \times 1.207 - \tfrac{1}{4} - \tfrac{1}{4} \leq 0, \text{ or } -\tfrac{1}{2} \leq 0.318 \leq \tfrac{1}{2} \qquad (29)$$

and hence QM does not violate the CH inequality in this possible world. In the case of the CHSH inequality, if all the counterfactual observables are zero, then also all the $\underline{\underline{E}} = 0$ in eq.25 and:

$$|E(\alpha,\beta) - E(\alpha,\beta')| + |E(\alpha',\beta') + E(\alpha',\beta)| \leq 8 \qquad (30)$$

or $S_{CHSH} \leq 8$, and hence QM does not violate the CHSH inequality in this possible world.

*4.D. The "QM-like" counterfactual values.*

An analyzer oriented at $\alpha$ and the same analyzer oriented at $\alpha'$ are incompatible situations in the classical way of thinking. In QM instead, an analyzer oriented at $\alpha$ has a probability $\cos^2(\alpha-\alpha')$ to behave as if it were oriented at $\alpha'$. It is therefore possible to measure at two different angle settings at the same time, with some probability (<1) for each alternative. Assuming then that the counterfactual $P_A(\alpha',t) = \cos^2(\alpha-\alpha') \times P_A(\alpha',t')$ (and the same for the other cases), the inequality in eq.6 becomes, for the usual angle settings:

$$-1 \leq \tfrac{1}{4} [\tfrac{1}{2} P_{AB}(\alpha,\beta') + P_{AB}(\alpha,\beta) + \tfrac{1}{2} P_{AB}(\alpha',\beta) + \tfrac{1}{4} P_{AB}(\alpha',\beta')] - \tfrac{1}{4}[P_{AB}(\alpha,\beta') + \tfrac{1}{2} P_{AB}(\alpha,\beta) + \tfrac{1}{4} P_{AB}(\alpha',\beta)$$
$$+ \tfrac{1}{2} P_{AB}(\alpha',\beta)] + \tfrac{1}{4}[\tfrac{1}{4} P_{AB}(\alpha,\beta') + \tfrac{1}{2} P_{AB}(\alpha,\beta) + P_{AB}(\alpha',\beta) + \tfrac{1}{2} P_{AB}(\alpha',\beta')] + \tfrac{1}{4} [\tfrac{1}{2} P_{AB}(\alpha,\beta') +$$
$$+ \tfrac{1}{4} P_{AB}(\alpha,\beta) + \tfrac{1}{2} P_{AB}(\alpha',\beta) + P_{AB}(\alpha',\beta')] - \tfrac{1}{4} [P_B(\beta) + \tfrac{1}{2} P_B(\beta)] - \tfrac{1}{4} [P_A(\alpha') + \tfrac{1}{2} P_A(\alpha')] \leq 0 \qquad (31)$$

and replacing the QM values for the probabilities:

$$-1 \leq 0.458 - \tfrac{3}{4} \leq 0, \text{ or: } -\tfrac{1}{4} \leq 0.458 \leq \tfrac{3}{4} \tag{32}$$

the CH bound is not violated, and therefore QM is not in contradiction with LR (nor with the "QM-like" possible world assumed). In the CHSH case, the value of (f.ex.) the observable A becomes $A(\alpha',t) = \cos^2(\alpha-\alpha') \times A(\alpha',t') = \tfrac{1}{2} \cdot A(\alpha',t')$ and the same for the other counterfactuals, what leads to $\underline{\underline{E(\alpha,\beta)}} = (5/4) \times E(\alpha,\beta)$ (recall that one term is a "double" counterfactual), so that:

$$(9/4)|E(\alpha,\beta) - E(\alpha,\beta')| + (9/4)|E(\alpha',\beta') + E(\alpha',\beta)| \leq 8 \tag{33}$$

therefore $S_{CHSH} \leq 32/9 \approx 3.55$, which is not violated by the QM predictions, and therefore QM is not in contradiction with LR (nor with the "QM-like" possible world assumed).

Summarizing this Section: depending on the (unavoidable) additional assumption or possible world chosen, the QM values may, or may not, violate the CH and CHSH inequalities.

**5. Discussion.**

The above presented results are disappointing: it is apparently impossible to solve the QM vs LR controversy by measuring the violation of a Bell's inequality. In what follows, I discuss some objections that have been raised to this conclusion.

*5.A "Time" is not an acceptable hidden variable.*

It is objected that the hidden variables are devised as internal degrees of freedom carried by the particles, and that "time" does not hold to this definition.

In response to this objection: it is possible to think of internal clocks carried by each particle, and then time becomes an internal variable. The truth in this debate is not clear to me. Faced to choose between LR and a debatable ban to time as a hidden variable, I find sensible to choose the latter as the false one. Besides, regardless of whether time is considered as a hidden variable or not, the time integrals in the eqs.16 and 21 do represent the way real measurements are performed.

*5.B The LR model is missing.*

It is objected that the argument in this paper is incomplete, for no model is presented explaining how the measured values violate the CH or CHSH bounds without violating LR. In other words: it is objected that, even if the logical support of the inference:

*violation of Bell's inequalities $\Rightarrow$ LR is false*

is weakened, the inference may be true anyway. To be sure that this inference is false, it is claimed, a model explaining (f.ex.) how eq.16 violates the CH bound without violating LR should be presented.

In response to this objection: note that if such a LR model were presented, the argument in this paper would be irrelevant. Such a model would provide a counterexample that would suffice to demonstrate the compatibility between LR and the violation of the Bell's inequalities. However, the claim in this paper is *not* that LR is true in the Nature, but just that the criterion we were using to settle experimentally the QM vs LR controversy was misleading. And, in consequence, that there is a logical basis to renew the search for LR models, now with a well definite constraint that may act as a useful guide (see the Summary).

*5.C Using time as a hidden variable may imply "action-at-a-distance".*

It is objected that time is treated here in a too naïve way, and that a consistent description should consider time according to Relativity.

In response to this objection: the different parts of the setup in the Fig.1 are static with respect to each other, so that a classical time, common to both stations, is well defined. Besides, the counterfactual situations have local causes. The counterfactuals involve the probability or expectation values for $A=\alpha$ when $A\neq\alpha$. The value of $\beta$ (or $\alpha$) is related with the probability or observable values at the station B (or A) only. The condition of Locality is always strictly valid.

*5.D Measurement independence is violated.*

It is objected that the presented argument introduces a correlation between the values of the hidden variable (time) and the values of the angle settings $\{\alpha,\beta\}$, therefore violating the premise of measurement independence. I find this objection most interesting.

In response to this objection: note that the correlation is not the hypothesis of some LR model, but the unavoidable consequence of the way real measurements are performed. Therefore, the meaning of the stated objection is that the premise of measurement independence is impossible to achieve in the real world. Bell's inequalities cannot be tested under the conditions holding to their premises, and hence there is no way they can be applied to experiments to solve the QM vs LR controversy. At this point, it is tempting to think that the "Fifth Position" conjectured by R.Gill, i.e., that the QM vs LR controversy is intrinsically undecidable [8], is perhaps correct.

However, I am not convinced that the stated objection necessarily has so devastating consequences. For, even if the correlation between time and angle settings is unavoidable, it is still possible to achieve a significant test by adding one specific assumption (f.ex., eq.17). In other words, it is true that:

*measurement independence valid* $\Rightarrow$ *Bell's inequalities are valid*

but it is still possible that Bell's inequalities are valid *even if* measurement independence is not valid. This is the case, f.ex., in the possible worlds described in the Sections 4.A and 4.B. For this reason, the condition of holding (or not) to the conditions defining these possible worlds becomes a useful tool to know if a given LR theory has been already refuted by the experiments (or not).

*5.E Comments.*

I guess that a specific property of time is the deepest reason of the weakening of the Bell's inequalities reported here. Time can take any given value only once. This property is what makes impossible to average over the space of the hidden variable for each angle setting, as it is required by the usual reasoning that leads to the Bell's inequalities. By the way, providing a satisfactory definition of time is an old an elusive problem. Perhaps, a working definition could be stated as: "time is the measurable magnitude that takes any given value only once". I do not find other measurable magnitude that holds to this definition.

It is also conceivable that the reported weakening is the consequence of a weak point in the usual way the values obtained from the theory of probability are linked to the observations. In the theory, events are thought to occur in abstract, independent parallel worlds. The average over an ensemble of these parallel worlds allows the simple calculation of probabilities. In any actual observation instead, events occur (and averages are obtained) successively in time, in the only available real world. The values obtained from these two different procedures are usually considered equal. But, these "ensemble" and "time" averages are not necessarily equal. The fact that they can be different has been recently applied to solve an old problem in the theory of gambling (St.Petersburg's) [9]. For the exploration of this interesting possibility, the discussions in [10-12] may provide relevant hints.

**Summary.**

The Bell's inequalities cannot be applied to measurements that occur during time (i.e., all real measurements, even in an ideally perfect setup) unless *at least one* additional and unavoidable assumption is made. This assumption means the definition of a "possible world" to calculate the numerical values of the counterfactual terms in the inequalities that can be deduced from LR *only*. In my opinion, the best choice to retrieve the usual Bell's inequalities is to assume the equality between the time averages of factual and counterfactual probabilities (or expectation values of observables). But, there is no fundamental reason why this equality must be true. It is reasonable then to conclude then that it is this equality, and not LR, what is disproved in the experiments reporting the violation of the Bell's inequalities.

In conclusion, the QM vs LR controversy cannot be settled in the way it has been expected. Nonetheless, the observed violation of the Bell's inequalities does restrict, in a significant way, the theories aimed to reconcile QM with LR. These theories must violate at least one of the additional assumptions or possible worlds where the Bell's inequalities are valid. F.ex., the ones described in the Sections 4.A and B. In this regard, note that 4A $\Rightarrow$ 4B, so that it suffices breaking 4B. In other words: the observed violation of the Bell's inequalities in the Nature implies that a LR theory can be acceptable *only if* the factual and counterfactual time averages predicted by such theory are different (note that the calculus of counterfactuals is a well defined operation in a LR theory). This is a simple criterion to decide if a LR model proposed in the future (say, one aimed to fulfill the claim in 5.B) has been already disproved by the experiments, or not.

**Acknowledgements.**

Many thanks to Mónica Agüero and Marcelo Kovalsky (this lab), Michael Hall (Australian National University, Canberra) and Peter Morgan (University of Yale, USA) for their helpful comments and suggestions.

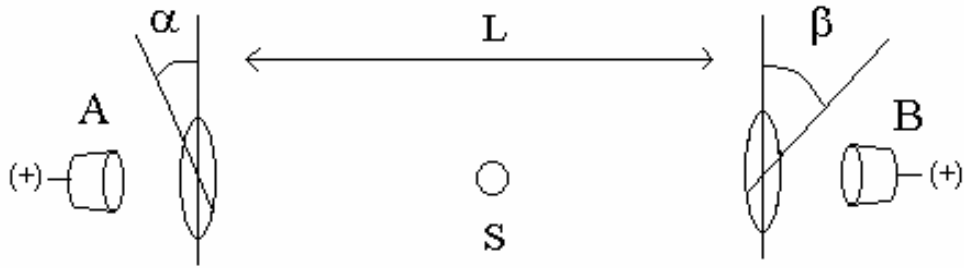

Figure 1: Scheme of a typical experiment measuring the Bell's inequalities. The source S emits pairs of photons entangled in polarization towards stations A and B separated by a large distance $L$. At each station, an analyzer is set at some angle. The probability of detection after both analyzers $P_{AB}(\alpha,\beta)$ is measured from the number of coincidences recorded in some time interval. Then the angle settings are changed and a new measurement is carried out. A detected photon is defined as a "+1" result. A non-detected photon (or a photon detected in the "reflected" output of the polarizer, not shown) is defined as a "-1" result.

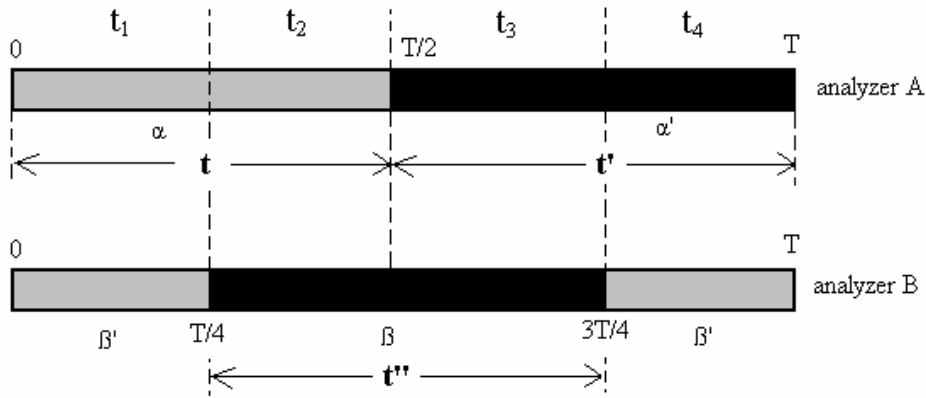

Figure 2: Distribution of the measuring time among the different angle settings for each analyzer, as it is assumed in the text. Time intervals in grey: the angles are $\alpha$ and $\beta$'; in black: $\alpha$' and $\beta$. This distribution is assumed for simplicity. Other possible distributions are equivalent to this one at the expense of a more involved notation. The different names of the variable "time" in each interval of integration are for the sake of clarity. It is assumed that the flux of pairs is constant in time.